\newcommand{\vecj}{\mbox{\boldmath$j$}}
\newcommand{\vecA}{\mbox{\boldmath$A$}}
\newcommand{\vecE}{\mbox{\boldmath$E$}}
\newcommand{\vecB}{\mbox{\boldmath$B$}}
\newcommand{\vecer}{\mbox{{\boldmath$\hat{r}$}}}
\newcommand{\vecnl}{\mbox{\boldmath$0$}}
\newcommand{\dfd}{{\rm d}}
\newcommand{\vecr}{\mbox{\boldmath$r$}}
\newcommand{\vecv}{\mbox{\boldmath$v$}}
\newcommand{\vecp}{\mbox{\boldmath$p$}}
\newcommand{\ie}{{\em i.e.}}
\newcommand{\eg}{{\em e.g.}}

\documentclass[aps,prl,twocolumn,groupedaddress]{revtex4}

\usepackage{graphicx}
\begin{document}


\title{On the nature of the plasma equilibrium}


\author{Hanno Ess\'en}
\email{hanno@mech.kth.se} \homepage{
http://www.mech.kth.se/~hanno/ }
\affiliation{Department of Mechanics, KTH\\ SE-100 44 Stockholm,
Sweden}


\date{2008 March}

\begin{abstract}
We calculate the energy of a homogeneous one component plasma
and find that the energy is lower for correlated motions of
the particles as compared to uncorrelated motion. Our starting
point is the conserved approximately relativistic (Darwin)
energy for a system of electromagnetically interacting
particles that arises from the neglect of radiation. For the
idealized model of a homogeneous one component plasma the
energy only depends on the particle canonical momenta and the
vector potential. The vector potential is then calculated in
terms of the canonical momenta using recent theoretical
advances and the plasma Hamiltonian is obtained. The result
can be understood either as due to the energy lowering caused
by the attraction of parallel currents or, alternatively, as
due to the inductive inertia associated with the flow of net
current.
\end{abstract}


\maketitle


Our theoretical understanding of matter is largely based on
the Coulomb  interaction between charged particles. For small
systems one can usually assume that the effects of the
magnetic corrections are secondary but for larger systems this
it not the case. Including magnetic interaction into the
theories, however, has met with considerable difficulties,
from the 1939 "magnetische Katastrophe" of Welker
\cite{welker2} to the 1999 rigorous proof of the instability
of matter with magnetic interaction by Griesemer and Tix
\cite{griesemer}. Their result is valid whether the magnetic
interaction is mediated by the (Darwin -) Breit potential or
via a quantized radiation field. In both these cases it is the
attraction of parallel currents that causes the problem. This
attraction, which is so fundamental that it is used in the
definition of the ampere, the unit of electric current in the
SI-system, indicates a long range energy lowering due to
correlation of currents that seems to diverge in many systems.

Laboratory and astrophysical plasmas are observed to harbor
intense currents  and magnetic fields \cite{BKkulsrud}, but in
plasma physics it is usually assumed that this is due to
non-equilibrium, and that the equilibrium plasma is described by
the traditional Maxwell-Boltzmann distribution (see \eg\ Burm
\cite{PLburm}). This is clearly at odds with the above findings of
an instability of the energy minimum ground state to parallel
current generation. Alastuey and Appel \cite{alastuey2} claim that
inclusion of the quantized radiation field removes the instability
problem, in direct contradiction with the findings of Griesemer
and Tix \cite{griesemer}. Most plasma physicists do not seem to be
aware of the problem even if there certainly has been a fair
amount of interest in energy minimizing states and self
organization, see \eg\ Woltjer \cite{woltjer}, Taylor
\cite{PLtaylor}.

Here we will show that within a simple standard model, based on
classical  electrodynamics and relativistic Hamiltonian mechanics,
the energy of a plasma is considerably reduced when the canonical
momenta are correlated and thus that conclusions drawn from the
traditional non-relativistic Maxwell-Boltzmann distribution of
non-interacting particles, or particles interacting only via a
Debye screened Coulomb potential, cannot be trusted. The model,
which neglects radiation, gives a quantitative estimate of the
energy reduction, but does not lead to any un-physical divergence.

Let us start from the following expression for the energy of a
system  of classical charged particles and electromagnetic fields,
\begin{equation}\label{eq.nonconstant.energy.kin.plus.em}
E = \sum_{a=1}^N \frac{m_a c^2}{\sqrt{1-\vecv_a^2/c^2}} +
\frac{1}{8\pi} \int (\vecE^2 + \vecB^2) \dfd V .
\end{equation}
If we think of the particles as confined to a finite volume one
finds that  energy leaves the volume due to electromagnetic
radiation (Poynting's theorem). If one neglects the radiation as a
higher order process ($\sim c^{-3}$) one finds that,
\begin{equation}\label{eq.constant.energy.kin.plus.em}
E_D = \sum_{a=1}^N \frac{m_a c^2}{\sqrt{1-\vecv_a^2/c^2}}  +
\frac{1}{2} \int (\phi\varrho + \frac{1}{c} \vecA \cdot \vecj)
\dfd V ,
\end{equation}
is a conserved energy \cite{essen99}. Inserting here that,
\begin{equation}\label{eq.curr.dens.points}
\varrho= \sum_{a=1}^N e_a \delta(\vecr -\vecr_a(t)),\;\; \vecj =
\sum_{a=1}^N e_a \vecv_a(t)\, \delta(\vecr -\vecr_a(t)) ,
\end{equation}
one obtains,
\begin{equation}\label{eq.constant.energy.kin.plus.em.rw}
E_D = \sum_{a=1}^N \left[ E_m(\vecv_a) +  \frac{e_a}{2}
\phi(\vecr_a) + \frac{e_a}{2c}\vecv_a \cdot \vecA(\vecr_a) \right]
,
\end{equation}
for the conserved (Darwin) energy, $E_D$, of a closed
non-radiating system of charged particles. Here $\phi$ and $\vecA$
are the scalar and vector potentials, in the Coulomb gauge, as
determined from the positions and velocities of the particles of
the system. These can be found from the approximately relativistic
Darwin approach
\cite{essen99,darwin,essen96,essen97,blazhyjevskyj&al}. From the
Darwin Lagrangian,
\begin{eqnarray}
\label{eq.LtotNoRad2} L_D = \sum_{a=1}^N \left[ L_m(\vecv_a) -
\frac{e_a}{2}\phi_a(\vecr_a) + \frac{e_a}{2c} \vecv_a \cdot
\vecA_a (\vecr_a)  \right] ,\\
\nonumber {\rm where,} \\
\label{eq.Lm} L_m(\vecv_a) = -m_a c^2\sqrt{1-\vecv_a^2/c^2} ,\\
\nonumber {\rm and,} \\
\label{eq.coul.pot} \phi_a (\vecr) = \sum_{b(\neq a)}^N
\frac{e_b}{|\vecr -\vecr_b|}, \\
\nonumber {\rm and,} \\
\label{eq.darwin.A.ito.velocity} \vecA_a (\vecr) = \sum_{b(\neq
a)}^N \frac{e_b [\vecv_b + (\vecv_b\cdot\vecer_b) \vecer_b]
}{2c|\vecr-\vecr_b|} ,\\
\nonumber {\rm where,} \\
\vecer_b=(\vecr -\vecr_b)/|\vecr -\vecr_b|,
\end{eqnarray}
one also finds the expression,
\begin{equation}\label{eq.mech.mom.expl.darwin.p.A}
\frac{m\vecv_a}{\sqrt{1-\vecv_a^2/c^2}} = \vecp_a - \frac{e_a}{c}
\vecA_a(\vecr_a),
\end{equation}
where, $\vecp_a = \partial L_D / \partial \vecv_a$, is the
canonical momentum.

Since the Darwin approach normally is considered accurate to order
$(v/c)^2$ it is sometimes stated that for consistency one must
also expand the relativistic particle energy expression, $E_m
(\vecv_a) =m_a c^2 /\sqrt{1-\vecv_a^2/c^2}$, to this order. There
are, however, several reasons not to do this. One is that whereas
such an expansion makes the particle energies qualitatively wrong
at high speeds there is no reason to assume that the other terms
in (\ref{eq.constant.energy.kin.plus.em.rw}) become qualitatively
wrong at high speeds \cite{blazhyjevskyj&al}. In fact one can show
\cite{essen07} that small modifications produce potentials that
are valid to arbitrary speeds. A second reason is that we are here
going for the Hamiltonian formalism and in this formalism speed is
not a relevant variable. The problems of finding the Hamiltonian
corresponding to the Darwin Lagrangian for large systems is
instead governed by a different dimensionless parameter, $N r_{\rm
e} /R$, where $N$ is the number of electrons, $r_{\rm e}$ is the
classical electron radius, and $R$ a typical length of the system
\cite{trubnikov2,jones&pytte,jones}.

We thus solve (\ref{eq.mech.mom.expl.darwin.p.A}) for the
velocity,
\begin{equation}\label{eq.v.over.c.ito.p.A}
\frac{\vecv_a}{c} =\frac{ \vecp_a - \frac{e_a}{c} \vecA_a }{\sqrt{
(m_a c)^2 +\left(\vecp_a - \frac{e_a}{c} \vecA_a \right)^2 }},
\end{equation}
and insert the result in
(\ref{eq.constant.energy.kin.plus.em.rw}). We note that here it is
assumed that the vector potential, $\vecA_a$, at particle $a$
excludes the contribution from that particle, as indicated in
(\ref{eq.darwin.A.ito.velocity}), so that self interaction is
avoided. Some algebra then gives the result,
\begin{eqnarray}\nonumber
E_D =\sum_{a=1}^N  m_a c^2  \frac{ 1+ \frac{\displaystyle
\left(\vecp_a - \frac{e_a}{c} \vecA_a \right)\cdot \left(\vecp_a -
\frac{e_a}{2c} \vecA_a \right)}{\displaystyle (m_a c)^2 } }{\sqrt{
1+ \frac{\displaystyle \left(\vecp_a - \frac{e_a}{c} \vecA_a
\right)^2}{\displaystyle (m_a c)^2 } }}  \\
\label{eq.darw.energy.p.A}\\  +\sum_{a=1}^N \frac{e_a}{2} \phi_a
 ,\nonumber
\end{eqnarray}
for the energy of our system of particles. This expression has
also been  derived directly by Legendre transformation from the
Darwin Lagrangian $L_D$ (Ess\'en \cite{essen96}).

The energy expression $E_D$ should be compared to the well known
energy  for particles in external fields,
\begin{equation}\label{eq.hamilt.part.syst.in.ext.fields}
E_A = \sum_{a=1}^N \left[ m_a c^2 \sqrt{1 +
\frac{\displaystyle\left(\vecp_a -\frac{e_a}{c}\vecA(\vecr_a)
\right)^2}{(m_a c)^2}}  + e_a \phi(\vecr_a) \right].
\end{equation}
The statistical mechanics of (the non-relativistic version of)
this Hamiltonian predicts zero magnetic response of a system of
classical charged particles according to the, so called, Bohr-van
Leeuwen theorem \cite{bohr,van_leeuwen,BKvanvleck}. The expression
(\ref{eq.darw.energy.p.A}) which is valid when the field is from
the particles themselves is different and the Bohr-van Leeuwen
theorem can not be invoked to make analogous predictions about its
properties.

In order to find a Hamiltonian from (\ref{eq.darw.energy.p.A}) we
must  express the vector potentials $\vecA_a$ in terms of the
canonical (generalized) momenta $\vecp_a$. In the Darwin approach
the vector potential can be obtained as solution of,
\begin{equation}\label{eq.vector.poisson.eq.t}
\nabla^2 \vecA = -\frac{4\pi}{c} \vecj_t ,
\end{equation}
where $\vecj_t$ is the transverse (divergence free) current
density \cite{BKjackson3}. One  can also use the ordinary current
density (\ref{eq.curr.dens.points}) and impose zero divergence by
a gauge transformation afterwards \cite{essen&nordmark}. In
equation (\ref{eq.vector.poisson.eq.t}) the second time derivative
in the d'Alembert operator of the wave equation has been skipped
since that term is $\sim c^{-2}$ and use of the Coulomb gauge
means that retardation is correctly handled to this order already
\cite{BKlandau2}. In order to get the vector potential in terms of
the $\vecp_a$ one might then insert,
\begin{equation}
\label{eq.veloc.resolved.one.c} \vecv_a =\frac{ \vecp_a}{m_a} -
\frac{e_a}{m_a c} \vecA_a (\vecr_a),
\end{equation}
from (\ref{eq.v.over.c.ito.p.A}) neglecting terms of order
$c^{-2}$ and higher, into the expression
(\ref{eq.curr.dens.points}) for the current density. For a
homogeneous one component plasma (with mobile particles of mass
$m$ and charge $e$) this approach transforms
(\ref{eq.vector.poisson.eq.t}) into,
\begin{equation}
\label{eq.no.approx.mom.space.poisson} \left( \nabla^2
-4\pi\frac{e^2}{mc^2} n(\vecr) \right) \vecA(\vecr) =
-\frac{4\pi}{c} \vecj_p(\vecr),
\end{equation}
where,
\begin{equation}\label{eq.numb.dens.points}
n(\vecr)= \sum_{a=1}^N  \delta(\vecr -\vecr_a),
\end{equation}
and [we are assuming $\vecj$ instead of $\vecj_t$ in
(\ref{eq.vector.poisson.eq.t})],
\begin{equation}
\label{eq.elec.momen.curr.dens} \vecj_p(\vecr) \equiv
 \frac{e}{m} \sum_{a=1}^N \vecp_a\,
\delta(\vecr-\vecr_a) ,
\end{equation}
essentially without approximation. Details are given in Es´s\'en
and Nordmark \cite{essen&nordmark}.

We now assume that we, at least for long range average purposes,
can replace the discontinuous number density
(\ref{eq.numb.dens.points}) with the smoothed volume average,
\begin{equation}\label{eq.numb.dens.vol.av}
\bar{n}(\vecr)=\frac{1}{V}\int_V \sum_{a=1}^N  \delta(\vecr
-\vecr_a) \dfd V .
\end{equation}
This, together with the assumption of a homogeneous, constant
density plasma, $\bar{n} =\,$constant, gives,
\begin{equation}
\label{eq.Lap.A.one.c.of.p.eq.yukawa} \left( \nabla^2
-\frac{1}{\lambda^2} \right) \vecA(\vecr) = -\frac{4\pi}{c}
\vecj_p(\vecr).
\end{equation}
Here,
\begin{equation}\label{eq.lambda.def}
\lambda^2 \equiv \frac{1}{4\pi \bar{n} r_{\rm e}},
\end{equation}
$r_{\rm e} \equiv \frac{e^2}{mc^2}$, and $\vecj_p$ is given by
(\ref{eq.elec.momen.curr.dens}). We assume, as usual, that the
homogeneous one component plasma consists of particles moving
against a background of smeared out charge of the opposite
sign of $e$, thus ensuring charge neutrality. The relevant
solution of Eq.\ (\ref{eq.Lap.A.one.c.of.p.eq.yukawa}) is
\cite{essen&nordmark},
\begin{equation}
\label{eq.yukawa.solut.one.c} \vecA_p(\vecr) = \frac{e}{mc}
\sum_{a=1}^N \frac{ \vecp_a \exp(-|\vecr-\vecr_a|/\lambda)}{
|\vecr-\vecr_a|},
\end{equation}
a Yukawa, or exponentially screened, vector potential. If we
can make this momentum space vector potential divergence free
we can use it in (\ref{eq.darw.energy.p.A}) to get the
Hamiltonian. We note that it has Coulomb singularities at the
particle positions, $\vecr=\vecr_a$, but that these vanish in
the Hamiltonian since there self interactions are assumed
excluded. Only the smooth part is thus of interest.

That exponential screening should occur in a Hamiltonian
describing magnetic interaction was noted by Bethe and Fr\"ohlich
\cite{bethe} already in 1933. They were not aware of the work of
Darwin, however. Neither were Bohm and Pines \cite{PLbohm} who,
studying collective motion of conduction electrons in 1951 also
arrived at such a result, directly from a particles plus field
Hamiltonian. The modern history of the exponential screening based
on the Darwin approach started in 1980 with Jones and Pytte
\cite{jones&pytte} who derived it in a Fourier transformed
formalism, also for a homogeneous one component plasma. Their
derivation was part of a debate on the usefulness of the original
Darwin Hamiltonian in plasma physics.

Noting that (\ref{eq.yukawa.solut.one.c}) is not divergence free
Ess\'en and Nordmark \cite{essen&nordmark} found the gauge
transformation to the correct Coulomb gauge divergence free
expression,
\begin{equation}
\label{eq.yukawa.solut.one.c.div.0} \vecA^c_p(\vecr) =
\frac{e}{mc} \frac{\exp(-r/\lambda)}{ r} [g(r/\lambda)\vecp +
h(r/\lambda)(\vecp \cdot\vecer)\vecer],
\end{equation}
for the vector potential from a particle with momentum $\vecp$ at
the origin. Here,
\begin{equation}
\label{eq.def.g.h} g(x)\equiv 1-\frac{\exp(x)-(1+x)}{x^2},\;\;
\mbox{and,}\;\; h(x)\equiv 2 - 3 g(x) .
\end{equation}
We now assume that all particles of the homogeneous one component
plasma of constant number density $\bar{n}$ have the same momenta
$\vecp$ and calculate the vector potential at the origin by
superposing contributions of the type
(\ref{eq.yukawa.solut.one.c.div.0}) all over space. It is
convenient to use spherical coordinates assuming that the momentum
$\vecp$ is in the $z$-direction. The contributions from the two
terms of (\ref{eq.yukawa.solut.one.c.div.0}) separately diverge so
it is necessary to do the angular integration of the second term
first. Because, $g(x)+h(x)/3=2/3$, their sum then gives us the
finite result,
\begin{equation}\label{eq.yukawa.vec.pot.const.dens.p.1}
\vecA^c_p = \vecA(\vecp) = \frac{2}{3} \frac{c }{e}\, \vecp .
\end{equation}
Note firstly that without the exponential damping such a vector
potential would diverge violently and secondly that the interplay
between the screening length $\lambda$ and the number density
$\bar{n}$ is such that both vanish from the final result of the
integration.

An alternative way of finding this result is by using Eq.\
(\ref{eq.Lap.A.one.c.of.p.eq.yukawa}) directly. We must then first
change the left hand side to the divergence free electric momentum
current density $\vecj_{pt}$. In the present case the current
density is constant, $\vecj_p(\vecr)=(e/m)\vecp\, \bar{n}$, so it
is not clear what the divergence free version should be. Realizing
that this current density is constant only because of volume
averaging, one can use a result by Crisp \cite{crisp} (in his
Appendix A) which states that the volume integral of a vector
field $\vecA$ is related to the volume integral of the
corresponding transverse field $\vecA_t$ by,
\begin{equation}\label{eq.crisps.formulas.for.long.and.trans.int.of.vf}
\int \vecA_t (\vecr) \dfd V = \frac{2}{3} \int \vecA (\vecr) \dfd V .
\end{equation}
When dealing with volume averages one should therefore take the
transverse part to be, $\vecj_{pt}=(2/3)\vecj_{p} = (2/3) (e/m)
\vecp\,\bar{n}$. Realizing that the corresponding, $\vecA^c_p
=\vecA_{pt}$, must then also be constant
(\ref{eq.Lap.A.one.c.of.p.eq.yukawa}) gives,
\begin{equation}
\label{eq.Lap.A.one.c.of.p.eq.yukawa.const} \left( 0
-\frac{1}{\lambda^2} \right) \vecA^c_p  = -\frac{4\pi}{c} \left(
\frac{2}{3}\frac{e}{m}  \vecp\, \bar{n} \right).
\end{equation}
Use of the expression (\ref{eq.lambda.def}) for $\lambda^2$ then
again gives (\ref{eq.yukawa.vec.pot.const.dens.p.1}), $\vecA^c_p =
(2/3) (c/e) \vecp$.

\begin{figure}
\centering
\rotatebox{0}{\includegraphics[width=240pt,height=190pt]{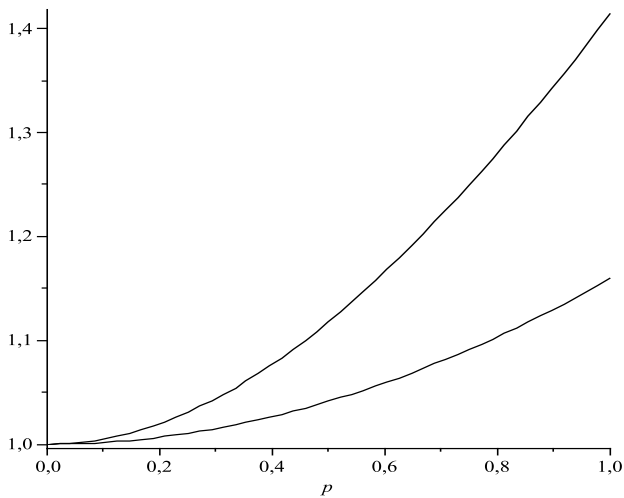}}
\protect\caption {The upper curve is the relativistic energy per
particle, $E_0(p)$ for free, uncorrelated, particles. The lower
curve is the Darwin energy $E_D (p)$, per particle, of a
homogeneous one component plasma with a constant momentum current
density, \ie\ with correlated momenta. The energy is in units of
$mc^2$ and the canonical momentum $p$ in units of $mc$.}
\label{darwin.energy.of.p.fig}
\end{figure}

Let us return to the energy (\ref{eq.darw.energy.p.A}) which, for
our case of a homogeneous one component plasma of particles, all
with the same momenta $\vecp$, gives,
\begin{equation}\label{eq.darw.energy.p.A.one.comp}
E_D (\vecp) = m c^2  \frac{ 1+ \frac{\displaystyle \left(\vecp -
\frac{e}{c} \vecA(\vecp) \right)\cdot \left(\vecp - \frac{e}{2c}
\vecA(\vecp) \right)}{\displaystyle (m c)^2 } }{\sqrt{ 1+
\frac{\displaystyle \left(\vecp - \frac{e}{c} \vecA(\vecp)
\right)^2}{\displaystyle (m c)^2 } }} ,
\end{equation}
{\it per} particle, if we ignore the contribution $e\phi_a /2$
from the scalar potential which averages to zero (and is
velocity and momentum independent in the Coulomb gauge). Here
we now insert the result
(\ref{eq.yukawa.vec.pot.const.dens.p.1}) and find after
simplification, that the Darwin energy {\it per} particle is,
\begin{equation}\label{eq.energy.homogeneous.plasm.const.curr}
E_D (p) =mc^2  \frac{ 1+\frac{2p^2}{9m^2c^2} }{
\sqrt{1+\frac{p^2}{9m^2c^2}}}.
\end{equation}
One should compare this result to the usual relativistic energy of
a free particle,
\begin{equation}\label{eq.rel.energy.free.part}
E_0 (p)= mc^2\sqrt{1+\frac{p^2}{m^2c^2}},
\end{equation}
which is obtained from (\ref{eq.darw.energy.p.A.one.comp}) when
$\vecA(\vecp) =\vecnl$, \ie\ for a non-magnetic plasma with
uncorrelated particle motions. These two functions are plotted in
Fig.\ \ref{darwin.energy.of.p.fig}. The connection between
velocity and momentum can be found from Eq.\
(\ref{eq.mech.mom.expl.darwin.p.A}). For our case, $p\, (1-2/3)  =
mv/\sqrt{1-v^2/c^2}$, so that $p=3mv/\sqrt{1-v^2/c^2}$. Inductive
inertia thus results in an effective  mass three times the normal
one.

The implications of the above result are far reaching for many
aspects of astrophysical, laboratory, and metallic conduction
electron plasmas.  The latter, which are unable to correlate their
momenta at higher temperatures, due to lattice oscillations,
become able to do so at lower temperatures. Their energy as
function of canonical momentum then shifts to the lower of the two
curves in Fig.\ \ref{darwin.energy.of.p.fig}. Inductive inertia
thus lowers the energy of the correlated electrons as originally
suggested by Frenkel \cite{frenkel}, see also Ess\'en
\cite{essen95}. For astrophysical plasmas it is not obvious
whether the nuclei or the electrons should constitute the fixed
background when making the one component plasma approximation. The
energy reduction is much greater if the heavy particles are
considered to be mobile. Kulsrud \cite{BKkulsrud} points out that
one can explain the longevity of astrophysical currents as due to
their large inductive inertia. Our result says that this inertia
in fact reduces the energy so that it also provides a mechanism
for the generation of these currents.

Recall that the approximations used to get the central result
(\ref{eq.energy.homogeneous.plasm.const.curr}) is (i) the neglect
of radiation, (ii) the neglect of terms of order $c^{-2}$, and
smaller in going from (\ref{eq.v.over.c.ito.p.A}) to
(\ref{eq.veloc.resolved.one.c}), and  (iii) in the use of the
smoothed density (\ref{eq.numb.dens.vol.av}) instead of the
discontinuous exact particle density.  These approximations seem
considerably less severe than those normally used in plasma
physics.


%








\end{document}